\documentclass[aip,
 mph,%
 amsmath,
 amssymb,
 superscriptaddress,
 floatfix,
 preprint,
 eqsecnum,%
 numerical,%
]{revtex4-1}

\usepackage{graphicx}
\DeclareGraphicsExtensions{.pdf,.png,.jpg}
\usepackage[caption=false
]{subfig}
\usepackage{epstopdf}
\usepackage{lipsum}
\usepackage{wrapfig}
\usepackage{wasysym}
\usepackage{threeparttable,booktabs,siunitx}
\usepackage{booktabs}
\usepackage{lmodern}
\usepackage{longtable}
\usepackage{multirow}
\usepackage{amsmath}
\usepackage[OT2,T1]{fontenc}
\usepackage[bookmarks=false]{hyperref}
\usepackage{color}
\hypersetup{pdfauthor={},pdfborder=0 0 0,colorlinks=true,citecolor=blue,linkcolor=blue,urlcolor=blue,}

\DeclareSymbolFont{cyrletters}{OT2}{wncyr}{m}{n}
\DeclareMathSymbol{\comb}{\mathalpha}{cyrletters}{"58}

\newcommand{\ben}{\begin{eqnarray}\displaystyle}
\newcommand{\een}{\end{eqnarray}}

\begin{document}

\title{Automatic image-domain Moir\'{e} artifact reduction method in grating-based x-ray interferometry imaging}

\author{Jianwei Chen}%
 \thanks{Jianwei Chen and Jiongtao Zhu have made equal contributions to this work and both are considered as the first authors.}
\affiliation{Paul C Lauterbur Research Center for Biomedical Imaging, Shenzhen Institutes of Advanced Technology, Chinese Academy of Sciences, Shenzhen, Guangdong 518055, People's Republic of China}
\affiliation{Department of Physics, Jinan University, Guangzhou, Guangdong 510632, People's Republic of China}
\author{Jiongtao Zhu}%
 \thanks{Jianwei Chen and Jiongtao Zhu have made equal contributions to this work and both are considered as the first authors.}
\affiliation{Paul C Lauterbur Research Center for Biomedical Imaging, Shenzhen Institutes of Advanced Technology, Chinese Academy of Sciences, Shenzhen, Guangdong 518055, People's Republic of China}
\affiliation{Department of Mechanical Engineering, Wuhan University of Technology, Wuhan, Hubei 430000, People's Republic of China}
\author{Wei Shi}%
\affiliation{Paul C Lauterbur Research Center for Biomedical Imaging, Shenzhen Institutes of Advanced Technology, Chinese Academy of Sciences, Shenzhen, Guangdong 518055, People's Republic of China}
\author{Qiyang Zhang}%
\affiliation{Paul C Lauterbur Research Center for Biomedical Imaging, Shenzhen Institutes of Advanced Technology, Chinese Academy of Sciences, Shenzhen, Guangdong 518055, People's Republic of China}
\affiliation{Research Center for Medical Artificial Intelligence, Shenzhen Institutes of Advanced Technology, Chinese Academy of Sciences, Shenzhen, Guangdong 518055, People's Republic of China}
\author{Hairong Zheng}%
\affiliation{Paul C Lauterbur Research Center for Biomedical Imaging, Shenzhen Institutes of Advanced Technology, Chinese Academy of Sciences, Shenzhen, Guangdong 518055, People's Republic of China}
\affiliation{Chinese Academy of Sciences Key Laboratory of Health Informatics, Shenzhen, Guangdong 518055, People's Republic of China}
\author{Dong Liang}%
 \thanks{Scientific correspondence should be addressed to Dong Liang (dong.liang@siat.ac.cn) and Yongshuai Ge (ys.ge@siat.ac.cn).}
\affiliation{Paul C Lauterbur Research Center for Biomedical Imaging, Shenzhen Institutes of Advanced Technology, Chinese Academy of Sciences, Shenzhen, Guangdong 518055, People's Republic of China}
\affiliation{Research Center for Medical Artificial Intelligence, Shenzhen Institutes of Advanced Technology, Chinese Academy of Sciences, Shenzhen, Guangdong 518055, People's Republic of China}
\affiliation{Chinese Academy of Sciences Key Laboratory of Health Informatics, Shenzhen, Guangdong 518055, People's Republic of China}
\author{Yongshuai Ge}%
 \thanks{Scientific correspondence should be addressed to Dong Liang (dong.liang@siat.ac.cn) and Yongshuai Ge (ys.ge@siat.ac.cn).}
\affiliation{Paul C Lauterbur Research Center for Biomedical Imaging, Shenzhen Institutes of Advanced Technology, Chinese Academy of Sciences, Shenzhen, Guangdong 518055, People's Republic of China}
\affiliation{Research Center for Medical Artificial Intelligence, Shenzhen Institutes of Advanced Technology, Chinese Academy of Sciences, Shenzhen, Guangdong 518055, People's Republic of China}
\affiliation{Chinese Academy of Sciences Key Laboratory of Health Informatics, Shenzhen, Guangdong 518055, People's Republic of China}

\date{\today}

\begin{abstract}
The aim of this study is to demonstrate the feasibility of removing the image Moir\'{e} artifacts caused by system inaccuracies in grating-based x-ray interferometry imaging system via convolutional neural network (CNN) technique. Instead of minimizing these inconsistencies between the acquired phase stepping data via certain optimized signal retrieval algorithms, our newly proposed CNN-based method reduces the Moir\'{e} artifacts in the image-domain via a learned image post-processing procedure. To ease the training data preparations, we propose to synthesize them with numerical natural images and experimentally obtained Moir\'{e} artifact-only-images. Moreover, a fast signal processing method has also been developed to generate the needed large number of high quality Moir\'{e} artifact-only images from finite number of acquired experimental phase stepping data. Experimental results show that the CNN method is able to remove Moir\'{e} artifacts effectively, while maintaining the signal accuracy and image resolution.
\end{abstract}

\maketitle

\section{Introduction}
The past two decades have witnessed the quick developments of grating-based x-ray interferometry imaging method, especially the Talbot-Lau imaging method. As a novel x-ray imaging method, it is able to generate three images, i.e., the absorption image, the differential phase contrast (DPC) image, and the dark-field (DF) image, with unique contrast mechanism from the same acquired dataset simultaneously. Different from the conventional absorption signal, the DPC signal corresponds to the x-ray refraction information. Studies have shown that the DPC signal may have advancements in providing superior contrast sensitivity for certain types of soft tissues\cite{Momose1996, Momose2006, Pfeiffer2006a, Bech2009, Jensen2011a, li2014grating}. In addition, the complimentary DF signal corresponds to the small-angle-scattering (SAS) information, and thus is particularly sensitive to certain fine structures such as microcalcifications inside breast tissue\cite{Anton2013, Michel2013, wang2014non, Grandl2015, scherer2016correspondence}. Therefore, numerous research interests\cite{Zhu2010, miao2013motionless, Ge2014, Koehler2015} have been attracted with the aim to translate this innovative x-ray imaging method from laboratory investigations to real clinical applications.

In an x-ray Talbot-Lau interferometry system, detection of the diffraction fringes of phase grating are often performed with an analyzer grating, which is usually an absorption grating having identical period as of the self-image of the phase grating. As a result, Moir\'{e} patterns with resolvable period by most of the medical grade flat-panel-detectors are generated. In reality, depending on the relative alignments of the phase grating and analyzer grating, the detected Moir\'{e} diffraction patterns may have various distributions, i.e., spatially varied periods and intensities, over the entire detector surface. To retrieve the attenuation, the refraction and the SAS information of an imaging object, the phase stepping technique\cite{Weitkamp2005} is usually utilized. Specifically, a group of Moir\'{e} patterns are recorded by laterally translating one of the three gratings within one period in Talbot-Lau interferometry. As a result, each detector pixel records a group of phase stepping data points, which form the so-called phase stepping curve and are used to extract the absorption, DPC, and DF signals. 

As long as all the experimental conditions are ideal, images free of artifacts can be easily generated. However, due to some non-ideal experimental settings, their image quality may be degraded. For example, the polychromatic x-ray beam used in laboratory may introduce the so-called type-II beam hardening effect\cite{bevins2013type} in DPC-CT imaging. For radiographic DPC imaging, the acquired DPC images may also be contaminated by phase wrapping artifacts\cite{jerjen2011reduction, tapfer2012experimental, epple2013unwrapping, epple2015phase} and Moir\'{e} image artifacts\cite{seifert2016optimisation, marschner2016helical, kaeppler2017improved, dittmann2018optimization, de2018analysis, Hauke2017Analytical, hauke2018enhanced}. The former artifact happens when the detected refraction angle of the imaging object is out of the $[0, 2\pi)$ range, while the second artifact usually happens due to the non-ideal experimental conditions. For example, the inaccurate mechanical movement of the grating during phase stepping procedure, the non-stationary drifts of the x-ray focal spot over multiple exposures, system vibrations, and so on. In this paper, the impact of these potential influences is categorized as effective phase stepping position deviation, denoted by $\eta$.

In fact, any slight phase stepping position deviation will break the consistency between the assumed ideal sinusoidal phase stepping signal model and the acquired phase stepping data. Due to such mismatch, undesired image artifacts might show up after signal extractions using the assumed theoretical signal model. Since this kind of artifacts look similar as of the Moir\'{e} diffraction patterns, so they are named as Moir\'{e} image artifacts. During experiments, these slight effective phase stepping position deviations are often random and hard to be exactly determined, therefore, Moir\'{e} image artifacts are often unavoidable and needs to be mitigated after data acquisitions. By far, several different methods have been proposed\cite{Pelzer2015Reconstruction, seifert2016optimisation, marschner2016helical, kaeppler2017improved, dittmann2018optimization, de2018analysis, hauke2018enhanced} to mitigate these Moir\'{e} image artifacts. Among them, Pelzer et al\cite{Pelzer2015Reconstruction} and Seifert et al\cite{seifert2016optimisation} have developed correction algorithms based on principle component analysis (PCA) to adjust the phase stepping position deviations. By establishing an expectation-maximization (EM) algorithm, the Moir\'{e} artifacts in single-shot DPC-CT imaging system could also be reduced\cite{marschner2016helical}. Kaeppler et al\cite{kaeppler2017improved} developed another phase step position optimization algorithm under the condition of discarding very irregular phase step data points. Moreover, an enhanced image reconstruction method\cite{hauke2018enhanced} has proposed to better fulfill a real clinical application environment. The phase step position deviations can also be corrected using a two-step fitting algorithm\cite{dittmann2018optimization}, in which both the sinusoidal shape of the phase stepping curve and the phase stepping positions are considered. Overall, most of the above Moir\'{e} artifacts reduction algorithms are developed with the aim to adjust the inaccurate phase stepping positions, namely, minimize the inconsistencies between the theoretical phase stepping signal model and the experimentally acquired phase stepping data during signal extractions.

Recently, deep learning technique\cite{lecun2015deep, goodfellow2016deep} has attracted huge research interests due to its remarkable performance in variety of applications, for example, image classification and segmentation, object identification, image artifact removal\cite{sun2018moire}, denoising, and so on. Taking advantages of its powerful strength and remarkable performance, it becomes very interesting to apply the deep learning technique to the Moir\'{e} artifacts reduction problem in x-ray interferometry imaging. In particular, we mainly consider the Moir\'{e} artifact reduction task as an image post-processing procedure, and hope to mitigate them in image-domain using an end-to-end supervised U-Net\cite{ronneberger2015u} type convolutional neural network (CNN). Unlike conventional algorithm-driven artifact reduction methods, the performance of CNN-based Moir\'{e} artifacts reduction method heavily relies on the training data, including both of the data quality and the data richness. In other words, sufficient number of high quality image pairs: one Moir\'{e} artifact image(known as network input) and one clean image (known as label), has to be prepared for network training. In this preliminary work, we suggest to train the network with numerical images, and validate its performance on experimental images. The numerical images used for network training are manually synthesized from natural images and experimentally acquired Moir\'{e} artifact-only images. Herein, we assume that all of the Moir\'{e} artifact-only images are acquired from an already fine-tuned Talbot-Lau interferometry with fixed grating alignments.

Even with same grating alignments, we noticed that the acquired Moir\'{e} artifacts are still different from each other, including the structural variations and the spatial shifts. To make the CNN network robust enough, undoubtedly, the Moir\'{e} artifact-only images used in numerical training data synthesis have to be sufficiently rich. In other words, repeated experiments have to be performed during training data preparations. This could dramatically prolong the data preparation period, and might become a hurdle when using the CNN technique. In addition, we also noticed that most of the Moir\'{e} artifact-only images generated from normal phase stepping data, which contains phase stepping uncertainties, and processed with standard signal extraction procedures have low signal-to-noise-ratio (SNR). Such low quality Moir\'{e} artifact-only image samples could also degrade the final CNN performance. Therefore, we proposed a fast signal processing method to generate large number of high SNR Moir\'{e} artifact only images from finite number of phase stepping datasets obtained with normal data acquisition procedures. The key idea is to manually increase the slight inconsistencies of the phase stepping data, thus, increasing the amplitude of the Moir\'{e} artifacts. Specifically, we suggest to manually replace one phase stepping image by its neighbors. For instance, replacing the $k$-$th$ phase stepping image by the $(k-1)$-$th$ or $(k+1)$-$th$ phase stepping image results in an increase of $\eta^{(k)}$ by one. By doing so, the inconsistencies between the theoretical signal model needed data and the experimental phase stepping data are greatly magnified. Results show that the amplitude of the Moir\'{e} artifacts can be dramatically improved. Moreover, roughly $M\times(M-1)$ number of Moir\'{e} artifacts with unique spatial distribution can be obtained from only one single experimental phase stepping data. As a result, the training data preparation procedure could be accelerated by at least ten times (assuming $M=4$).

The remain of this paper is organized as following: the section II presents the theoretical analyses of the phase stepping position deviation induced Moir\'{e} artifact for the absorption, DPC, and DF contrasts. In the section III, numerical simulations are performed to demonstrate the proposed novel signal processing method. In the section IV, we describe the details of generating the training dataset, and the proposed image-domain Moir\'{e} artifact reduction CNN network. In section V, numerical experiments are carried out to verify the performance of the CNN-based Moir\'{e} artifact reduction method. In addition, experimental results are also validated. We made discussions about this work in section VI, and finally gave a brief conclusion in section VII.

\section{Theoretical analyses}
\subsection{General theory of Moir\'{e} artifacts}
\begin{figure*}[t]
\centering
\includegraphics[width=0.70\textwidth]{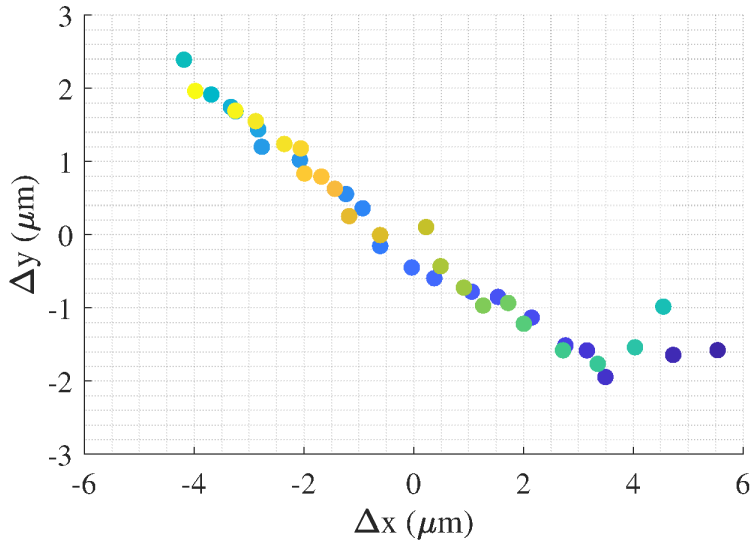}
\vspace{-0.2in}
\caption{Experimentally measured x-ray tube focal spot drift map. The x-axis is along the horizontal direction, and y-axis is along the vertical direction. On average, the spot drift along x-axis between two consecutive exposures is $0.46\pm0.04\; \mu m$, and $0.20\pm0.02\; \mu m$ along y-axis.}
\label{fig:spotdrift} 
\end{figure*}

In this theoretical discussion part, the standard phase stepping procedure is assumed. With this ideal signal model, the detected x-ray intensity at a certain pixel for the $k$-th phase stepping position is denoted as
\begin{equation}
{I}^{(k)}=I_0\Bigg\{1+\epsilon\cos\left[\frac{2\pi k}{\mathrm{M}}+\phi\right]\Bigg\},
\label{eq:signal_mod}
\end{equation}
where $I_0$ corresponds to the x-ray absorption contrast signal, $\phi$ corresponds to the x-ray differential phase contrast signal, and $\epsilon$ corresponds to the x-ray dark-field contrast signal, $\mathrm{M} (\ge3)$  corresponds to the total phase step number, and $k=1,2,...,\mathrm{M}$. Notice that $I_0$, $\epsilon$, and $\phi$ are all varied spatially. No Poisson photon fluctuations are assumed in this model. To proceed, the effective phase stepping position deviation $\eta^{(k)}$ is considered. In reality, $\eta^{(k)}$ could bee caused by many external influences, vibrations, and mechanical inaccuracies. For our experimental system, we noticed that $\eta^{(k)}$ are dominantly influenced by the x-ray tube focal spot drifts between consecutive exposures, see Fig.~\ref{fig:spotdrift}. Now the acquired x-ray intensity can be written as:
\begin{align}
\label{eq:signal_1}
{I}^{(k)}=&I_0\Bigg\{1+\epsilon\cos\left[2\pi\frac{k+\eta^{(k)}}{\mathrm{M}}+\phi\right]\Bigg\}.
\end{align}
Without knowing $\eta^{(k)}$ precisely in prior, the $I_0$, $I_1$, and $\phi$ signals should still be retrieved using Eqs.~(\ref{eq:ch01_N0})-(\ref{eq:ch01_phi}) from these acquired non-ideal phase stepping datasets $\{{I}^{(k)}\}$. With Taylor approximations similar as in literature\cite{Hauke2017Analytical}, the final radiographic absorption image $\hat{\mathcal{A}}$ with Moir\'{e} artifacts can be estimated via
\begin{align}
\hat{\mathcal{A}}=-\ln \left\{ \frac{\hat{I^{obj}_0}}{\hat{I^{ref}_0}}\right\} &\approx \mathcal{A}-\frac{2\pi \epsilon_{ref}}{\mathrm{M}^2}\sum_{k=1}^{\mathrm{M}}\eta^{(k)}_{ref}\sin\left[\frac{2\pi k}{\mathrm{M}}+\phi_{ref}\right] \nonumber \\
&+\frac{2\pi \epsilon_{obj}}{\mathrm{M}^2}\sum_{k=1}^{\mathrm{M}}\eta^{(k)}_{obj}\sin\left[\frac{2\pi k}{\mathrm{M}}+\phi_{obj}\right].
\label{eq:atten_0}
\end{align}
Herein, the superscript $ref$ denotes the air scan, and the superscript $obj$ denotes the object scan. During these derivations, we have ignored the high order terms of $\alpha^{(k)}_{ref}$, $\alpha^{(k)}_{obj}$, $\eta^{(k)}_{ref}$, and $\eta^{(k)}_{obj}$. Similarly, the radiographic DF contrast image with Moir\'{e} artifacts, denoted as $\hat{\mathcal{S}}$, can be expressed as below
\begin{align}
\hat{\mathcal{S}}=-\ln \left\{ \frac{\hat{\epsilon^{obj}}}{\hat{\epsilon^{ref}}}\right\}& \approx \mathcal{S}-\frac{2\pi}{\mathrm{M}^2}\sum_{k=1}^{\mathrm{M}}\eta^{(k)}_{ref}\sin\left[\frac{4\pi k}{\mathrm{M}}+2\phi_{ref}\right] \nonumber \\
& +\frac{2\pi}{\mathrm{M}^2}\sum_{k=1}^{\mathrm{M}}\eta^{(k)}_{obj}\sin\left[\frac{4\pi k}{\mathrm{M}}+2\phi_{obj}\right] \nonumber \\
& +\frac{2\pi \epsilon_{ref}}{\mathrm{M}^2}\sum_{k=1}^{\mathrm{M}}\eta^{(k)}_{ref}\sin\left[\frac{2\pi k}{\mathrm{M}}+\phi_{ref}\right] \nonumber \\
& -\frac{2\pi \epsilon_{obj}}{\mathrm{M}^2}\sum_{k=1}^{\mathrm{M}}\eta^{(k)}_{obj}\sin\left[\frac{2\pi k}{\mathrm{M}}+\phi_{obj}\right].
\label{eq:sas_0}
\end{align}
Finally, the DPC image, denoted as $\hat{\mathcal{P}}$, can be expressed by
\begin{align}
\hat{\mathcal{P}}=\hat{\phi^{obj}}-\hat{\phi^{ref}}&\approx \mathcal{P} + \frac{2\pi}{\mathrm{M}^2}\sum_{k=1}^{\mathrm{M}}\eta^{(k)}_{ref}\cos\left[\frac{4\pi k}{\mathrm{M}}+2\phi_{ref}\right] \nonumber \\
&-\frac{2\pi}{\mathrm{M}^2}\sum_{k=1}^{\mathrm{M}}\eta^{(k)}_{obj}\cos\left[\frac{4\pi k}{\mathrm{M}}+2\phi_{obj}\right].
\label{eq:dpc_0}
\end{align}
In Eqs.~(\ref{eq:atten_0})-(\ref{eq:dpc_0}), the $\mathcal{A}$, $\mathcal{S}$, and $\mathcal{P}$ denote the ideal absorption image, DF image, and DPC image free of Moir\'{e} artifacts, correspondingly. Obviously, the measured absorption image $\hat{\mathcal{A}}$, DF image $\hat{\mathcal{S}}$, and DPC image $\hat{\mathcal{P}}$ are all contaminated by Moir\'{e} artifacts due to the random $\eta^{(k)}_{ref}$ and $\eta^{(k)}_{obj}$. In particular, the residual Moir\'{e} artifact on absorption image $\hat{\mathcal{A}}$ only contains the basic frequency component, which is identical as of the detected Moir\'{e} patterns from the phase stepping procedure. The residual Moir\'{e} artifacts on DPC images $\hat{\mathcal{P}}$ only consists of doubled frequency component. Whereas, Moir\'{e} artifacts on DF images $\hat{\mathcal{S}}$ are superimposed by the basic and doubled frequency components. Since the $\epsilon$ factor is usually less than one, therefore, the doubled frequency component becomes more dominant than the basic frequency component, see Eq.~(\ref{eq:sas_0}). Due to the same reason, Moir\'{e} artifacts on absorption images are less pronounced than on DPC and DF images. Additionally, theoretical results also indicate that large values of $\eta^{(k)}_{ref}$ and $\eta^{(k)}_{obj}$ can lead to more dramatic Moir\'{e} artifact amplitude. Approximately, the Moir\'{e} artifacts can be assumed as additive signals\cite{de2018analysis} to the artifact-free images if ignoring the influences of $\epsilon_{ref}$, $\epsilon_{ref}$, object refractive property $\mathcal{P}$, and object SAS property $\epsilon_{obj}$. Although not quite rigorous, such assumptions typically yield good results during experimental validations.

\subsection{Numerical simulations}
Numerical simulations are performed in Matlab (The MathWorks Inc., Natick, MA, USA) to quantitatively investigate the impact of $\eta^{(k)}$ to the Moir\'{e} artifact magnitude. First of all, $I^{obj}_0$ and $I^{obj}_1$ are assumed to be the same as $I^{ref}_0$ and $I^{ref}_1$, which are set to 100 and 20, respectively. The $\phi^{ref}=\mathrm{mod}\left(\sqrt{((u/5+50)^2+(v/5+50)^2)}, 2\pi\right)$ map varies along both the horizontal and vertical directions, denoted as $u$ and $v$. Moreover, the $\eta^{(k)}_{ref}$ and $\eta^{(k)}_{obj}$ are sampled randomly from two random distributions: the uniform random distribution $[-\sigma_{\eta}, \sigma_{\eta}]$, and the Gaussian distribution with zero mean and standard deviation of $\sigma_{\eta}$ ($\sigma_{\eta}\in[0.0, 0.3]$). The phase shift of the object $\mathcal{P} = \phi^{obj}-\phi^{ref}$ is fixed at $\pi/3$. The total number for phase steps is equal to eight. The simulated images have dimension of 256 pixels by 256 pixels, and no Poisson photon fluctuations are considered. Finally, simulations with same parameters are repeated by 50 times.

\begin{figure*}[t]
\centering
\includegraphics[width=1\textwidth]{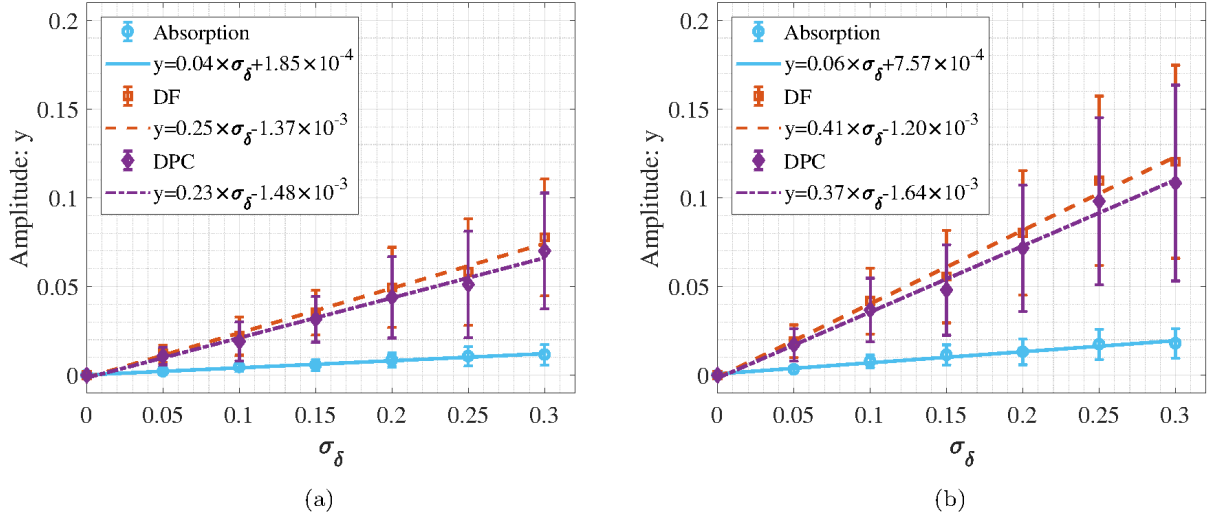}
\vspace{-0.4in}
\caption{Plots of Moir\'{e} artifact amplitude: (a) uniform random distribution $[-\sigma_{\eta}, \sigma_{\eta}]$, the coefficient of determinations, denoted $R^{2}$, of the linear fittings are 0.9899, 0.9930, and 0.9896 for the absorption, DF, and DPC, respectively. (b) Gaussian distribution with standard deviation of $\sigma_{\eta}$, the $R^{2}$ of the linear fittings are 0.9680, 0.9745, and 0.9652 for the absorption, DF, and DPC, respectively.}
\label{fig:Num_results_2} 
\end{figure*}

Results in Fig.~\ref{fig:Num_results_2} clearly demonstrate that the amplitude of the Moir\'{e} artifacts is linearly proportional to the effective phase stepping position deviation intensity $\sigma_{\eta}$. Therefore, it is always beneficial for generating images with less Moir\'{e} artifacts by minimizing the effective phase stepping position deviations. Although the phase stepping position deviation strengths need to be well controlled when designing a real experimental system, however, we would like to artificially increase their strengths to generate Moir\'{e} artifact-only images with higher SNR. Details of this method are discussed in the next section.

\section{Moir\'{e} artifact reduction CNN}
\subsection{Training data preparations}
As demonstrated by the numerical simulations, generally, larger phase step position deviations correspond to higher Moir\'{e} artifact amplitude for all the three different contrast mechanisms, see Fig.~\ref{fig:Num_results_2}. Based on this observation, we were inspired to develop a fast signal processing method to generate sufficient number of CNN network needed high SNR Moir\'{e} artifact-only image samples from a standard x-ray Talbot-Lau imaging system, from which phase stepping datasets are acquired with standard radiation dose levels. To this aim, we suggest to virtually increase the deviation $\eta^{(k)}$ strength at the $k$$-th$ phase stepping position. In particular, one of the initially acquired $M$ phase stepping images is replaced by the image corresponding to other phase step position. Clearly, such manual replacement operation explicitly increases the strength of $\eta^{(k)}$ (i.e., $\sigma^{k}_{\eta}$), and thus helps increasing the inconsistencies of the phase stepping dataset. For instance, if replacing the second phase step image by the third phase step image, then $\eta^{(k=2)}$ approximately enlarges by 1.

\begin{figure*}[t]
\centering
\includegraphics[width=0.8\textwidth]{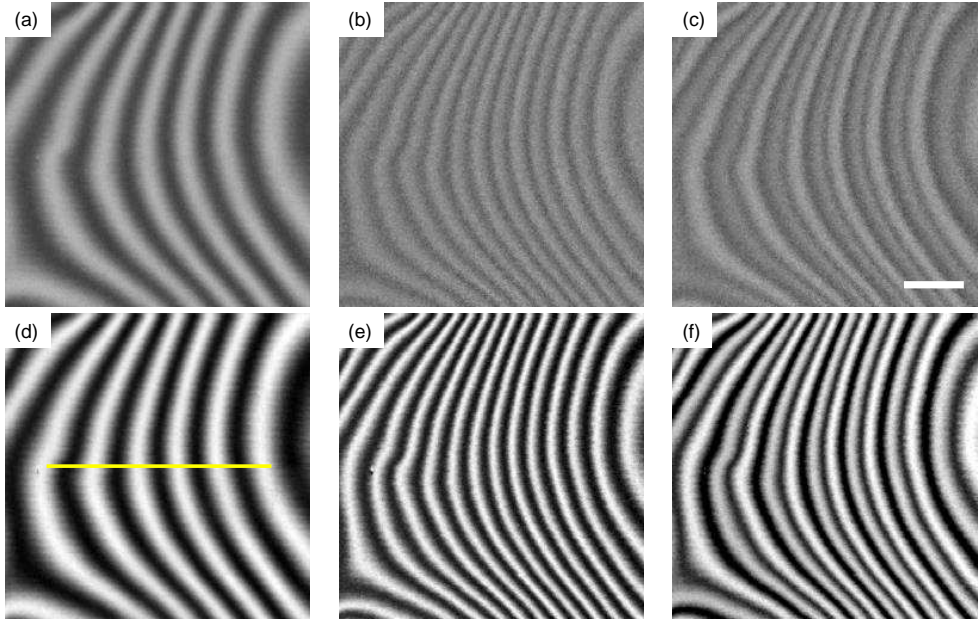}
\vspace{-0.15in}
\caption{The experimentally generated Moir\'{e}-artifact-only images from two different air-scan phase stepping datasets. In particular, images in (a)-(c) are obtained from standard data processing (denoted as Original), while images in (d)-(f) are obtained by replacing the second phase step image with the third one in reference data, and the third phase step image with the fourth one in object scan (denoted as Processed). From the left column to the right column, they are for absorption (display window: [-0.14, 0.16]), DPC (display window: [-0.47, 0.46]), and DF (display window: [-0.47, 0.46]), correspondingly. The scale bar denotes 1.0 cm.}
\label{fig:Exp_moire_0} 
\end{figure*}

\begin{figure*}[t]
\centering
\includegraphics[width=1.0\textwidth]{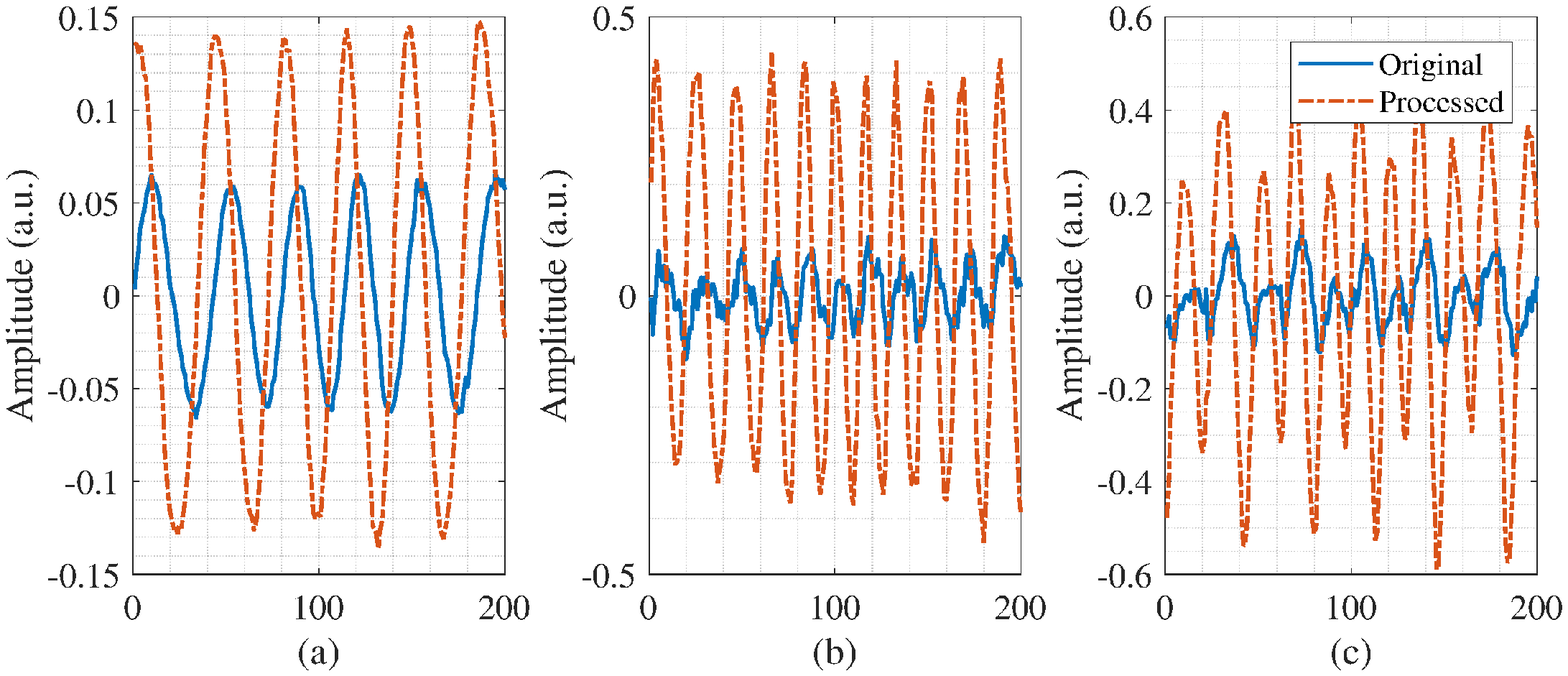}
\vspace{-0.5in}
\caption{Line profile results of central row of Moir\'{e}-artifact-only images in Fig.~\ref{fig:Exp_moire_0}:  (a) is for absorption results, (b) is for DPC results, (c) is for DF results. The solid lines correspond to the standard data processing method, and the dashed lines correspond to the proposed data processing method.}
\label{fig:Exp_moire_1} 
\end{figure*}

As shown in Fig.~\ref{fig:Exp_moire_0} and Fig.~\ref{fig:Exp_moire_1}, the resultant amplitudes of the Moir\'{e} artifacts indeed are increased significantly. In this study, we generate the Moir\'{e} artifact-only images by only replacing one phase stepping image once at a time. Line profiles in Fig.~\ref{fig:Exp_moire_1}(b)-(c) demonstrate that this new signal processing method can also reduce image noise. As a result, the quality of the Moir\'{e} artifact-only images can be greatly enhanced. Meanwhile, since there are $M$ phase stepping images, and multiple air-scan phase stepping datasets can be easily acquired during system warm up or calibration scans, therefore, the network required large number of Moir\'{e} artifact-only image samples can be quickly generated.


Afterwards, the network training data are ready to be synthesized. In this work, 5100 natural images(downloaded from ImageNet) and 210 experimentally acquired Moir\'{e}-artifact-only images (amplitude normalized between 0.0 to 1.0) with high SNR are prepared for each contrast mechanism, denoted as $\mathcal{M}_{\mathcal{A}}$, $\mathcal{M}_{\mathcal{S}}$, and $\mathcal{M}_{\mathcal{P}}$, see Fig.~\ref{fig:Exp_moire_0}(j)-(l). First, the central part of the image with dimension of $256\times256$ is cropped and converted to gray-scale image $\mathrm{I}_{n}$. Afterwards, images are normalized and combined accordingly with the experimentally acquired Moir\'{e} artifact only images to generate the needed network training dataset for each contrast mechanism: the absorption image $\mathcal{A}_{Moir\'{e}}$, the dark-field image $\mathcal{S}_{Moir\'{e}}$, and the differential phase contrast image $\mathcal{P}_{Moir\'{e}}$. Details of these operations are expressed below:
\begin{align}
\label{eq:M0}
\mathcal{A}_{Moir\'{e}} &= (3-\alpha) \times\frac{\mathrm{I}_{n}}{2^{8}}+\alpha\times\mathcal{M}_{\mathcal{A}}, \\
\label{eq:M1}
\mathcal{S}_{Moir\'{e}} &= (3-\alpha) \times\frac{\mathrm{I}_{n}}{2^{8}}+\alpha\times\mathcal{M}_{\mathcal{S}}, \\
\label{eq:M2}
\mathcal{P}_{Moir\'{e}} &=(2\pi-\alpha) \times\frac{\partial}{\partial u}\left[\frac{\mathrm{I}_{n}}{2^{8}}\right]+\alpha\times\mathcal{M}_{\mathcal{P}}.
\end{align}
These Moir\'{e} samples are randomly augmented by factor $\alpha$, which is uniformly sampled between 0.0 and 0.3. Notice that the selection of 0.3 is very empirical. For every individual contrast mechanism, 200 out of the 210 experimentally acquired Moir\'{e}-artifact-only images are randomly sampled to synthesize the 5000 pairs of network training dataset, and the rest 10 experimentally acquired Moir\'{e}-artifact-only images are used to generate the 100 pairs of network testing dataset.

\begin{figure*}[t]
\centering
\includegraphics[width=1\textwidth]{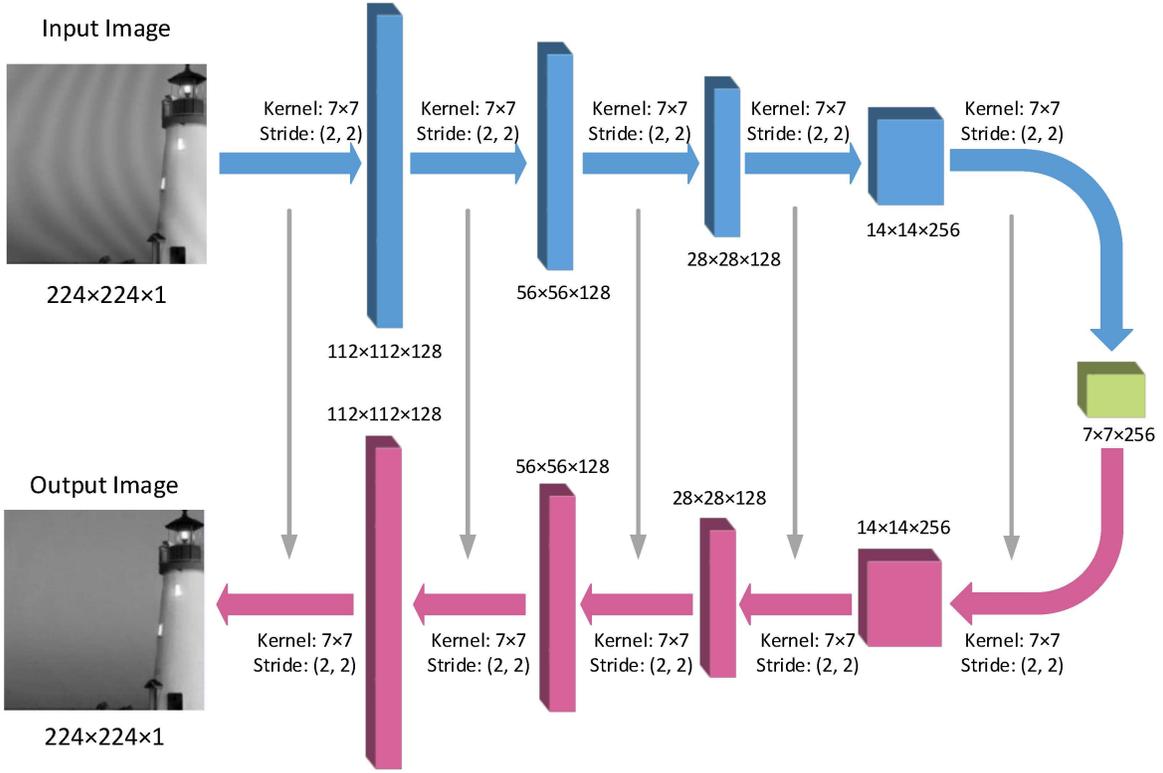}
\vspace{-0.4in}
\caption{The architecture of the U-Net type CNN network used in this study. Five shortcuts (gray vertical arrows) are connected symmetrically between the encoder and the decoder.}
\label{fig:network_ed} 
\end{figure*}
\subsection{CNN network}
The U-Net type CNN network architecture is implemented, as illustrated in Figure~\ref{fig:network_ed}. It contains ten layers in total. The strides of the $7\times7$ sized convolutional filters are set to be (2, 2). Five shortcuts are added accordingly to minimize the vanishing gradient phenomenon during back-propagation\cite{he2016deep}. The activation functions are ReLu\cite{hahnloser2000digital} (Rectified Linear Unit). Using the same CNN network, the synthesized absorption, DF, and DPC dataset are trained separately. For each network training, the objective loss function of the network is defined as:
\begin{equation}
\mathcal{L} = \frac{1}{\mathrm{N_u}}\frac{1}{\mathrm{N_v}}\sum^{\mathrm{N_u}}_{u=1}\sum^{\mathrm{N_v}}_{v=1}|\mathrm{I_m}-\mathrm{I^{cnn}_m}|^2,
\label{eq:loss}
\end{equation}
where $\mathrm{N_u}$ and $\mathrm{N_v}$ represent the total pixel number of the image along the $u$ and $v$ directions, correspondingly. In addition, the reference image free of Moir\'{e} artifact is denoted as $\mathrm{I_m}$, the CNN-learned image with reduced Moir\'{e} artifact is denoted as $\mathrm{I^{cnn}_m}$. During training procedures, the Adam algorithm\cite{kingma2014adam} is used with learning rate of 0.0001. The network is trained for 1000 epochs with mini-batch size of 64, and batch-shuffling is turned on to increase the randomness of the training data. On the Tensorflow deep learning framework using a single NVIDIA GeForce GTX 1070 GPU, the training time for each contrast is around eight hours.

\section{Results of numerical experiments}
\begin{figure*}[t]
\centering
\includegraphics[width=1\textwidth]{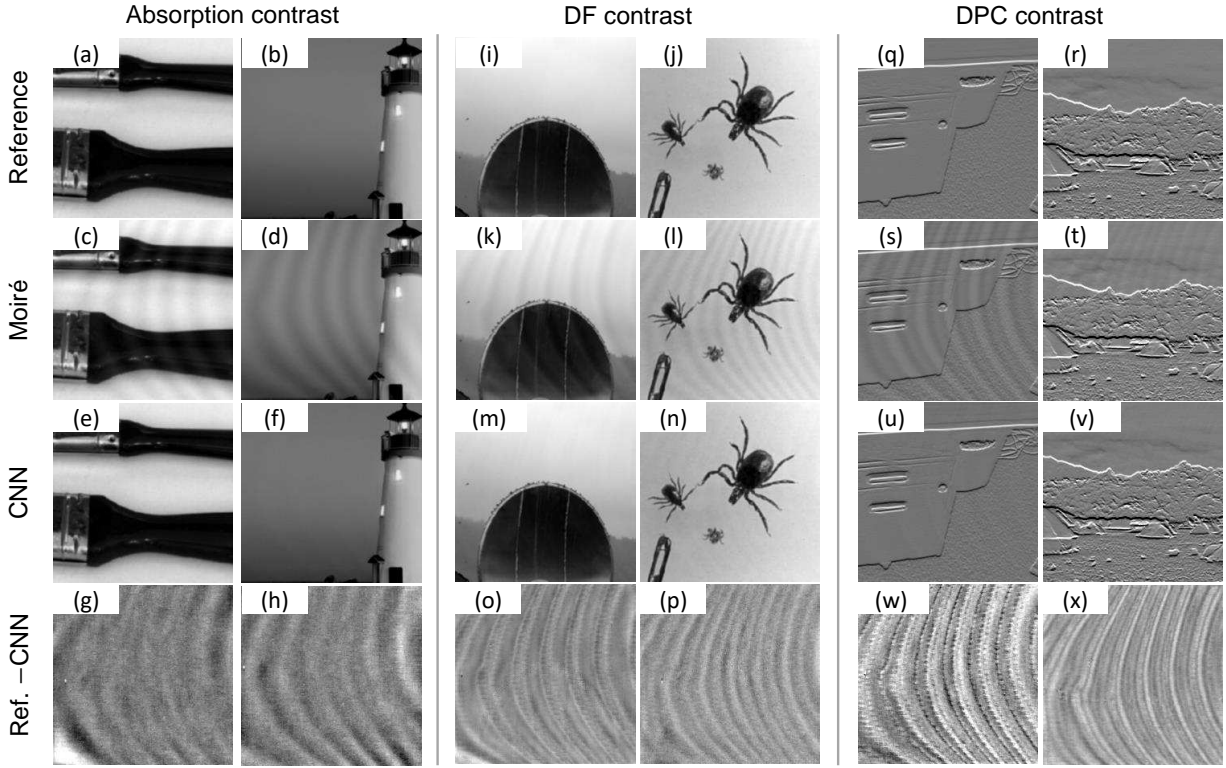}
\caption{The numerical validation results for absorption, DF, and DPC images. The display window for absorption and DF images is $[0.00, 3.00]$, and for DPC images is $[0, 2\pi]$. The display window for difference images in the bottom row is $[-0.10, 0.10]$.}
\label{fig:Num_cnn_1} 
\end{figure*}
\subsection{Numerical validation results of the CNN}
The network validation results in Figs.~\ref{fig:Num_cnn_1} demonstrate that the proposed CNN is robust enough to reduce the Moir\'{e} artifacts for the absorption contrast, DF contrast, and DPC contrast. The difference images between the reference images and the CNN learned images with reduced Moir\'{e} artifacts show that this method barely degrade the image spatial resolution. Despite that some very minor Moir\'{e} artifacts may still existed after processed by the CNN, they are visually negligible on the CNN processed results. In addition, the CNN-based artifact mitigation method does not introduce any new image artifacts.

To quantitatively characterize the performance of the CNN network, the structural similarity (SSIM) index are calculated for the entire image of three different contrast mechanisms independently. Herein, the reference image is considered as the ground truth. Comparisons results are listed in Table.~\ref{table:1}. Clearly, the CNN is able to significantly improve the image SSIM index, demonstrating that most of the Moir\'{e} artifacts have been removed after processed by the CNN.
\begin{table*}[h]
\centering
\caption{The structural similarity (SSIM) index values of the entire image (reference image is considered as the ground truth) before and after processed by the CNN network for three different contrast mechanisms.}
\vspace{0.5em}

\label{table:1}
  \begin{tabular}{cccc}
    \hline
    {} &
  \multicolumn{1}{c}{\;\;\;Absorption contrast\;\;\;} &
  \multicolumn{1}{c}{\;\;\;Dark-field contrast\;\;\;} &
  \multicolumn{1}{c}{\;\;\;Differential phase contrast\;\;\;} \\[2pt]
   \hline
    Before CNN  & 0.919$\pm$0.049 & 0.811$\pm$0.063 &  0.877$\pm$0.091\\[3pt]

    After CNN & 0.994$\pm$0.004 &  0.963$\pm$0.013 & 0.975$\pm$0.025\\
    \hline
  \end{tabular}
\end{table*}

\section{Experiments and results}
\subsection{Experiments}
Experiments are performed on an in-house x-ray Talbot-Lau interferometry imaging bench in our lab. The system includes a rotating-anode Tungsten target diagnostic-level X-ray tube (IAE XM15, IAE S.p.A., ITALY). It is operated at 40.00 kV (with mean energy of 28.00 keV) radiographic mode with 0.10 mm nominal focal spot. The X-ray tube current is set at 100.00 mA, with a 3.00 second exposure period for each phase step. The X-ray detector is an energy-integrating detector (Varex 3030Dx, Varex Imaging Corporation, UT, USA) with a native element dimension of 194.00 $\mu$m $\times$ 194.00 $\mu$m. The Talbot-Lau interferometry consists of three gratings: a source grating G0, a $\pi$-phase grating G1, and an analyzer grating G2. The G0 grating has a period of 24.00 $\mu m$, with a duty cycle of 0.35. The G1 grating has a period of 4.36 $\mu m$, with a duty cycle of 0.50. The G2 grating has a period of 2.40 $\mu m$, with a duty cycle of 0.50. The distance between G0 and G1 is 1773.60 mm, and the distance between G1 and G2 is 177.36 mm. The total phase stepping number $\mathrm{M}$ of the G0 grating is set to eight, with a stepping interval of 3.00 $\mu m$. The linear stage (VP-25XL, Newport, USA) has a high precision of 0.01 $\mu m$. Two specimens, a chicken claw and a chicken drumstick, are scanned separately with two different dose levels: the reference 100\% dose and a lower dose level of 15\%.

To determine the focal spot drifts of our x-ray tube, two tungsten bids of diameter 1.0 $mm$ are scanned. They are taped on a rigid steel holder, which is fixed on the optical table. The bids are positioned close to the G0 grating, which is located 15.00 cm downstream of the source. In total, 40 independent exposures are collected from two individual experiments. The second is performed one week after the first, but with exactly the same tube settings as of the DPC data acquisitions. With the acquired bids absorption images and the imaging geometry, the tube focal spot drift map is estimated, see Fig.~\ref{fig:spotdrift}.

To collect the Moir\'{e}-artifact-only sample images $\mathcal{M}_{\mathcal{A}}$, $\mathcal{M}_{\mathcal{S}}$, and $\mathcal{M}_{\mathcal{P}}$ in Eqs.~(\ref{eq:M0})-(\ref{eq:M2}), phase stepping datasets without any object are acquired repeatedly for four times at the reference dose level. Using the above discussed data preparation method, 210 high SNR Moir\'{e} artifact-only image samples (dimension of $256\times256$) with varied spatial distributions are generated for each contrast mechanism.

\subsection{Experimental results}
\begin{figure*}[t]
\centering
\includegraphics[width=1\textwidth]{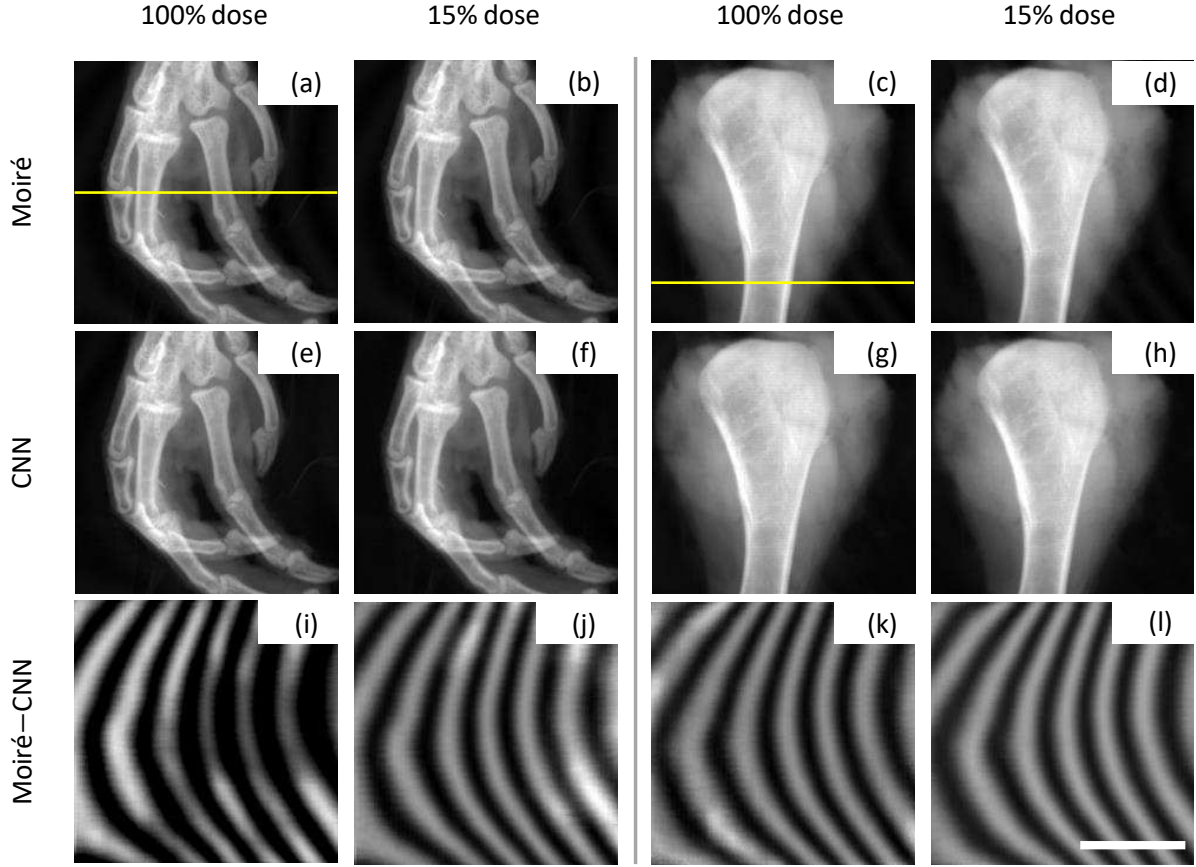}
\caption{The experimental absorption imaging results. The display window for the difference images is [-0.10, 0.10]. The scale bar denotes 2.00 cm.}
\label{fig:Exp_results_1} 
\end{figure*}

\begin{figure*}[t]
\centering
\includegraphics[width=1\textwidth]{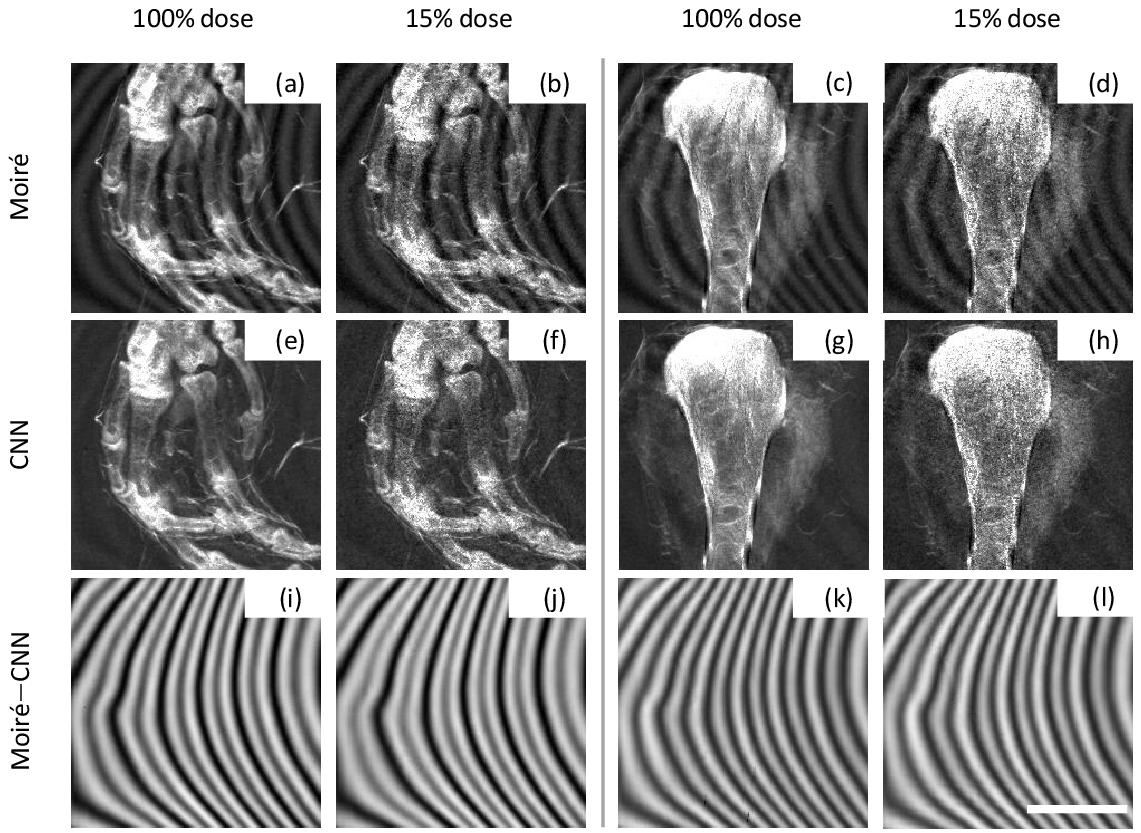}
\caption{The experimental DF imaging results. The display window for the difference images is [-0.20, 0.20]. The scale bar denotes 2.00 cm.}
\label{fig:Exp_results_2} 
\end{figure*}

\begin{figure*}[t]
\centering
\includegraphics[width=1\textwidth]{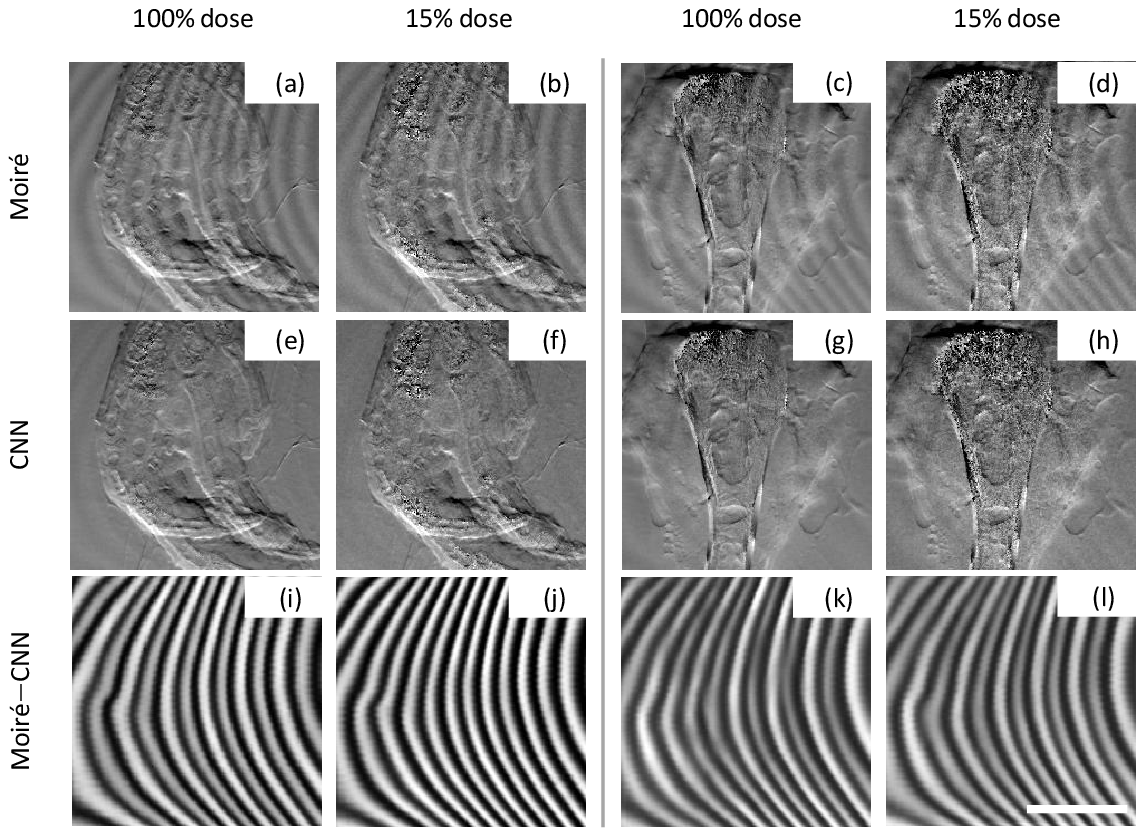}
\caption{The experimental DPC imaging results. The display window for the difference images is [-0.20, 0.20]. The scale bar denotes 2.00 cm.}
\label{fig:Exp_results_3} 
\end{figure*}

\begin{figure*}[t]
\centering
\vspace{-0.5in}
\includegraphics[width=0.90\textwidth]{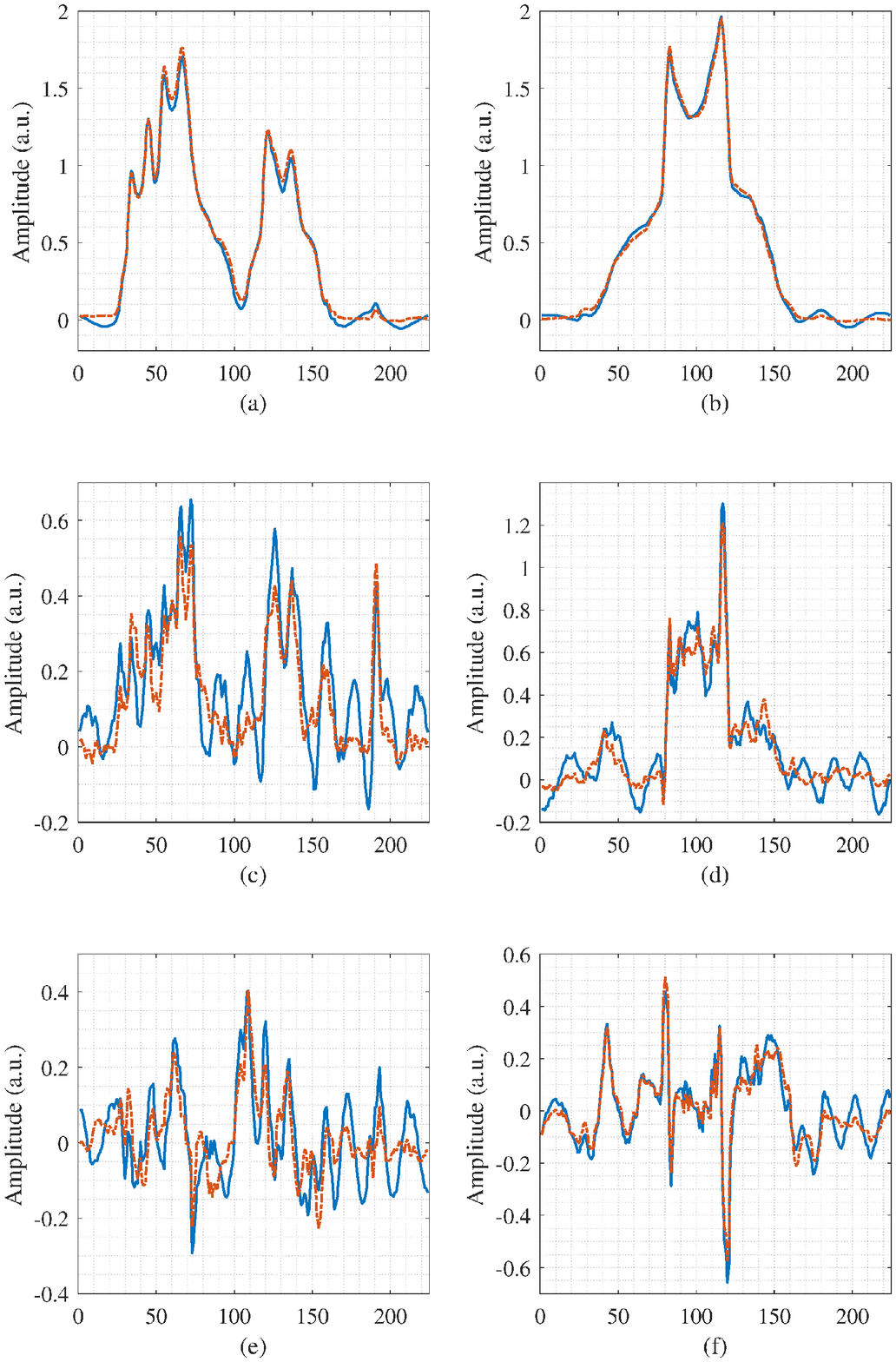}
\vspace{-0.8in}
\caption{Line profile of the experimental results obtained with $100\%$  radiation dose level: (a)-(b) are for absorption images, (c)-(d) are for DF images, (e)-(f) are for DPC images. The left column corresponds to the chicken claw specimen, and the right column corresponds to the chicken drum stick specimen. Please refer the lines on Fig.~\ref{fig:Exp_results_1}(a) and (c) for certain locations. Herein, the solid lines are drawn from original images before CNN is processed, while the dashed lines are drawn from images after CNN is processed.}
\label{fig:Exp_results_4} 
\end{figure*}

The Moir\'{e} artifact reduction results of the CNN network using experimentally acquired phase stepping datasets are shown in Figs.~\ref{fig:Exp_results_1}-\ref{fig:Exp_results_3} separately for the absorption contrast, DF contrast, and DPC contrast. Undoubtedly, results demonstrate that the already trained CNNs are able to remove most of the Moir\'{e} artifacts for the three different contrast mechanisms, while maintaining high image spatial resolution and signal accuracy, see plots in Fig.~\ref{fig:Exp_results_4}. As discussed previously, Moir\'{e} artifacts on the acquired absorption images are way less pronounced than on the DF and DPC images, however, the CNN can still effectively remove them. Results also demonstrate that the CNN network is also feasible to reduce Moir\'{e} artifacts on images acquired with lower radiation dose levels. Therefore, the image quality can be significantly improved after processed by the proposed CNN method.

\section{Discussion}
In this work, we have demonstrated a CNN enabled image-domain Moir\'{e} artifacts reduction method for grating-based x-ray interferometry imaging system. This method considers the Moir\'{e} artifacts reduction task as a image post-processing procedure, rather than an optimization task of signal extractions from phase stepping datasets. Therefore, the conventional signal retrieval method is still utilized to extract the three contrast images. Afterwards, these extracted images are feed into the proposed CNN network to get rid of the Moir\'{e} artifacts and finally generate clean images. Experimental results show that the CNN method can effectively reduce image Moir\'{e} artifacts, while maintaining the signal accuracy and image resolution.

The network training data with Moir\'{e} artifacts are synthesized from natural images and the experimentally acquired Moir\'{e} artifact-only images. We believe this might be the easiest way to generate the needed training data, which always requires a large volume of high quality of training samples. To do so, we have made two assumptions: The first one assumes that Moir\'{e} artifacts are independent of the scanned object and thus can be approximated as additive signals to the artifact-free images. The second one assumes that the generated Moir\'{e} artifact-only samples from finite number of experiments are rich enough to represent all of the possible Moir\'{e} artifacts appear on real data acquired from the same interferometry imaging system with identical grating alignments. Although not rigorous enough, experimental results demonstrate that the first assumption is able to yield good results. During our experimental validations, we did not find violations to the second assumption. However, the performance of the CNN could be degraded if Moir\'{e} artifacts appear on real data are very different from the collected Moir\'{e} artifact-only samples. This might be considered as one limitation of our currently proposed method. Whenever this happens, the new Moir\'{e} artifact patterns have to be added into the Moir\'{e} artifact-only sample pool. Afterwards, the CNN networks have to be retrained with newly synthesized training data. Sometimes, new network architecture with deeper dimension\cite{goodfellow2014generative, huang2017densely, sun2018moire} may be beneficial.

Although majority of the Moir\'{e} artifacts on absorption image, DPC image, and DF image can be effectively mitigated, some minor artifacts may still remain after processed by the CNN. Two main reasons may cause this result. First of all, the used Moir\'{e}-artifact-only samples in this work only contain 200 images. This may not be able to represent all of the possible Moir\'{e} artifacts generated from experiments. If increasing the number of Moir\'{e}-artifact-only samples in future, the performance of the CNN could be further optimized. The second reason may be related to the used U-Net type CNN network. In this work, it is purely an empirical choice. Other types of CNN architectures\cite{goodfellow2014generative, huang2017densely, sun2018moire} with more deep network depth or more network parameters may generate better Moir\'{e} artifacts reduction results.

\section{Conclusion}
In conclusion, we have demonstrated a CNN-based image-domain Moir\'{e} artifacts reduction method for grating-based x-ray interferometry imaging system. The new method treats the Moir\'{e} artifact reduction task as an image-domain post-processing procedure, and the CNN is utilized as an efficient tool to realize the aim. In addition, a fast signal processing method to generate large number of high quality Moir\'{e} artifact-only image samples has also been demonstrated to ease the training data preparations. Experimental results show that the CNN is able to effectively remove the Moir\'{e} artifacts while maintaining the signal accuracy and image resolution.

\section*{Acknowledgment}
The authors would like to thank Dr. Peiping Zhu at the Institute of High Energy Physics, Chinese Academy of Sciences, for his patient discussions to the theoretical analyses results. The authors also would like to acknowledge the ImageNet organization for providing the natural images to enable the network training procedures. This project is supported by the Young Scientists Fund of the National Natural Science Foundation of China (Grant No.~11804356).

\appendix
\section{}
Analytically, the three unknown signals can all be estimated from Eq.~(\ref{eq:signal_mod}) and Eq.~(\ref{eq:I1_signal}) as follows:
\begin{align}
\label{eq:ch01_N0}
\hat{I_0} =&  \frac{1}{\mathrm{M}} \sum_{k=1}^{\mathrm{M}} {I}^{(k)}, \\
\label{eq:ch01_N1} 
\hat{I_1} =&  \frac{2}{\mathrm{M}} \sqrt{\left[\sum_{k=1}^{\mathrm{M}} {I}^{(k)}\sin(\frac{2\pi k}{\mathrm{M}})\right]^2+ \left[\sum_{k=1}^{\mathrm{M}}{I}^{(k)}\cos(\frac{2\pi k}{M})\right]^2}, \\
\label{eq:ch01_phi}
\hat{\phi} =& \tan^{-1}\left[- \frac{\sum_{k=1}^{\mathrm{M}} {I}^{(k)}\sin(\frac{2\pi k}{\mathrm{M}})}{\sum_{k=1}^{\mathrm{M}} {I}^{(k)}\cos(\frac{2\pi k}{\mathrm{M}})} \right].
\end{align}
After adding the random fluctuation term to each phase stepping position $k$, the modulation term in Eq.~(\ref{eq:signal_1}) can be approximated by ignoring the high order terms of the Taylor expansion with respect to the small value $\eta^{(k)}$, namely,
\begin{align}
&\cos\left[2\pi \frac{k+\eta^{(k)}}{\mathrm{M}}+\phi\right] \nonumber \\
=&\cos\left[\frac{2\pi k}{\mathrm{M}}+\phi\right]\cos\left[\frac{2\pi \eta^{(k)}}{\mathrm{M}}\right] - \sin\left[\frac{2\pi k}{\mathrm{M}}+\phi\right]\sin\left[\frac{2\pi \eta^{(k)}}{\mathrm{M}}\right], \nonumber \\
\approx&\cos\left[\frac{2\pi k}{\mathrm{M}}+\phi\right] - \frac{2\pi \eta^{(k)}}{\mathrm{M}}\sin\left[\frac{2\pi k}{\mathrm{M}}+\phi\right].
\label{eq:app_1}
\end{align}
As a result,
\begin{align}
&\sum_{k=1}^{\mathrm{M}} {I}^{(k)}\sin\left[\frac{2\pi k}{\mathrm{M}}\right]\approx -\frac{\mathrm{M}I_1}{2}\sin\left[\phi\right]\nonumber \\
& - \frac{\pi I_1}{\mathrm{M}}\sum_{k=1}^{\mathrm{M}}\eta^{(k)}\left[\cos\left[\phi\right]-\cos\left[\frac{4\pi k}{\mathrm{M}}+\phi\right]\right].
\label{eq:app_2}
\end{align}
Similarly, we have
\begin{align}
&\sum_{k=1}^{\mathrm{M}} {I}^{(k)}\cos\left[\frac{2\pi k}{\mathrm{M}}\right]\approx \frac{\mathrm{M}I_1}{2}\cos\left[\phi\right] \nonumber \\
& - \frac{\pi I_1}{\mathrm{M}}\sum_{k=1}^{\mathrm{M}}\eta^{(k)}\left[\sin\left[\phi\right]+\sin\left[\frac{4\pi k}{\mathrm{M}}+\phi\right]\right].
\label{eq:app_3}
\end{align}

\bibliography{Bibliography_Paper}

\end{document}